\documentclass[10pt,superscriptaddress,prl,aps,longbibliography,twocolumn,floatfix]{revtex4-2}
\usepackage[T1]{fontenc}
\usepackage{lmodern}
\usepackage{amsmath,amssymb,amsthm,mathtools,bm,graphicx,microtype,booktabs}
\usepackage{xcolor}
\usepackage{algpseudocode}

\usepackage[colorlinks=true,linkcolor=magenta,citecolor=teal,urlcolor=teal]{hyperref}
\newtheoremstyle{letterresult}{4pt}{4pt}{\itshape}{}{\bfseries}{.}{.4em}{}
\theoremstyle{letterresult}

\newcommand{\ket}[1]{|#1\rangle}

\newcommand{\FNL}{F_{\rm NL}}

\graphicspath{{figures/}{./}}
\begin{document}

\title{The magical memory cost of entanglement spectrum compression}

\newcommand{\Aalto}{Department of Applied Physics, Aalto University, FI-00076 Aalto, Espoo, Finland}
\newcommand{\ICMM}{Interdisciplinary Centre for Mathematical Modelling and Department of Mathematical Sciences,\\ Loughborough University, Loughborough, Leicestershire LE11 3TU, United Kingdom}
\newcommand{\LUPhys}{Department of Physics, Loughborough University, Loughborough, LE11 3TU, United Kingdom}
\newcommand{\Tampere}{Computational Physics Laboratory, Physics Unit, Faculty of Engineering and Natural Sciences, Tampere University, P.O. Box 692, FI-33014 Tampere, Finland}
\newcommand{\Helsinki}{Helsinki Institute of Physics P.O. Box 64, FI-00014, Finland}

\author{Matias Karjula}
\affiliation{\Aalto}
\author{Vilma Nieminen}
\affiliation{\Aalto}
\author{Teemu Ojanen}
\email[\vspace{-3pt}]{teemu.ojanen@tuni.fi}
\affiliation{\Tampere}
\affiliation{\Helsinki}
\author{Tapio Ala-Nissila}
\email[\vspace{-3pt}]{tapio.ala-nissila@aalto.fi}
\affiliation{\Aalto}
\affiliation{\ICMM}
\author{Moein N. Ivaki}
\email[\vspace{-3pt}]{moein.najafiivaki@aalto.fi}
\affiliation{\Aalto}

\begin{abstract}
How much quantum memory is needed to faithfully retain the entanglement spectrum of a bipartite pure state after separating out Bell pairs through local unitaries? We derive the exact tradeoff between Bell-pair yield, residual memory, and reconstruction fidelity under local unitaries, defining an operational measure of entanglement-spectrum complexity. Smoothed entanglement spread bounds this memory cost from below, while diverging nonlocal magic forces unbounded memory at any fixed positive fidelity. Chaotic circuit dynamics exhibit a transient memory peak whose height increases with system size, followed by declining cost despite continued entropy growth. A large typical Haar-random state requires only an accuracy-dependent memory, whereas stationary fermionic Gaussian states retain a growing cost with size. In the circuits studied, the multiscale spectral structure associated with entanglement catalysis also has a practical consequence for matrix product state compression, yielding a multiplicative overhead not captured by the conventional entropy-based bond dimension estimate.
\end{abstract}

\maketitle
\textbf{\emph{Introduction.}} 
Random quantum dynamics can generate volume-law entanglement from product states~\cite{fisher2023random,Nahum2017}, raising questions about both the resources needed to represent the resulting states and the entanglement transformations they support. These are closely related through the Schmidt probabilities, the eigenvalues of a reduced density matrix that form the entanglement spectrum~\cite{Horodecki2009,PhysRevLett.101.010504}. Their cumulative weight determines the bond dimension required for a matrix product state approximation across a cut at a prescribed fidelity~\cite{PhysRevLett.93.040502,PhysRevLett.100.030504,SCHOLLWOCK201196}, while their distribution constrains how entanglement can be manipulated by local operations~\cite{ChitambarGour2019,LipkaBartosik2024,Nielsen1999,Vidal1999,vanDamHayden2003,PhysRevLett.83.3566}. For many identical copies, entropy sets the leading rates of quantum source coding and entanglement concentration into Bell pairs~\cite{jozsa1994new,Schumacher1995,Bennett1996}, while fluctuations of the Schmidt weights introduce additional resource requirements in coherent concentration and dilution~\cite{LoPopescu1999,HaydenWinter2003,Harrow2010,HarrowShor2010}.
This motivates asking how much of a single state's entanglement can be coherently compressed into Bell pairs, and how much residual quantum memory must be retained for faithful reconstruction.

For a single copy, local unitaries preserve every Schmidt probability. Bell pairs contribute a uniform factor to the spectrum, so an exact decomposition into Bell pairs and a residual state must retain all spectral nonuniformity in that state. In particular, $b$ Bell pairs have Schmidt rank $2^b$ and entanglement entropy $S_1=b$, yet require no residual. At a fixed reconstruction accuracy, the minimum memory required for such coherent compression thus provides an operational measure of entanglement-spectrum complexity. This residual memory is closely related to the notion of nonlocal magic~\cite{Cao2025Gravity,PRXQuantum.6.020324}. While nonstabilizerness~\cite{gottesman1998heisenberg} is studied as a resource in quantum information and computing~\cite{PhysRevA.71.022316,Veitch2014,HowardCampbell2017,Leone2022}, and as a way to characterize many-body states and dynamics~\cite{White2021,LiuWinter2022,minar2026magic,odavic2025stabilizer,ebner2026global,karjula2026ebbs}, minimizing it over local changes of basis isolates the contribution associated with shared entanglement. Bipartite pure stabilizer states provide a zero-memory reference, as local Clifford unitaries transform them into Bell pairs and local product states~\cite{Fattal2004}. For general pure states, the fidelity-based measure of nonlocal magic is determined entirely by the entanglement spectrum~\cite{Sierant2026,Viscardi2026}. Recent studies of random quantum dynamics have further connected near scale-free entanglement spectra to diverging nonlocal magic and universal entanglement embezzlement~\cite{Karjula2026,Aditya2026,zhang2026entanglementembz}, which allows any fixed entangled target to be borrowed while returning the shared catalyst with vanishing disturbance~\cite{vanDamHayden2003,Zanoni2024}. These findings further motivate a quantitative formulation of the relation between catalytic capabilities and spectral compressibility.

Here we derive the exact tradeoff between Bell-pair yield, residual memory, and reconstruction fidelity for arbitrary bipartite pure states under unrestricted local unitaries. We show that diverging nonlocal magic forces unbounded residual memory at any fixed positive fidelity, and smoothed entanglement spread provides an accuracy-dependent lower bound. Simulations of chaotic circuits reveal a transient memory peak that recedes while entanglement entropy continues to rise. Around this peak and at fixed fidelity, the near multiscale spectral structure imposes a multiplicative overhead beyond the entropy-based bond dimension estimate \(2^{S_1}\). In contrast, a large Haar-typical state asymptotically admits an accuracy-dependent, size-independent remainder, whereas stationary fermionic Gaussian states retain a cost that grows with size. In particular, every universally embezzling family requires unbounded residual memory at any fixed positive fidelity, establishing an obstruction to representing such families through Bell pairs and a bounded residual.

\textbf{\emph{Coherent entanglement spectrum compression.}} 
Consider a (single copy of) pure bipartite state \(\ket\psi_{AB}\) of \(N=2n\) qubits, with \(n\) qubits held by each party \(A\) and \(B\). Our goal is to faithfully compress the state using local unitaries by separating its maximally entangled Bell-pair structure from a residual register that carries the remaining nonflat entanglement spectrum. We can write the Schmidt decomposition as \(\ket\psi_{AB}=\sum_{i=0}^{2^n-1}\sqrt{\lambda_i}\ket{a_i}_A\ket{b_i}_B\), with \(\lambda_0\ge\lambda_1\ge\cdots\ge0\) and zero padding when the Schmidt rank is smaller than \(2^n\). For an allocation \((b,k)\) with \(b+k\le n\), we divide each party's \(n\) qubits into a Bell register of \(b\) qubits, a residual of \(k\) qubits, and \(r=n-b-k\) qubits in a fixed pure product state. The desired factorized approximation is
\begin{align}
 &(\hat U_A\otimes \hat U_B)\ket\psi_{AB}
 \approx\ket{\Omega_{b,k}},\nonumber\\
 &\ket{\Omega_{b,k}}:=
 \ket{\Phi_2}^{\otimes b}_{A_b B_b}\otimes
 \ket{\gamma_k}_{A_k B_k}\otimes
 \ket0^{\otimes r}_{A_r}\ket0^{\otimes r}_{B_r},
 \label{eq:target}
\end{align}
where $\ket{\Phi_2}:=(\ket{00}+\ket{11})/\sqrt2$ is a symmetric Bell state, and $\ket{\gamma_k}$ is an arbitrary normalized bipartite pure state. More generally, \(|\Phi_2\rangle\) may be replaced by any maximally entangled two-qubit state. The state-dependent unitary encoders \(\hat U_{A,B}\) act on their entire subsystems with unrestricted complexity. With a sufficient initial supply of Bell pairs, the same representation may also be seen as a preparation protocol~\cite{Bennett1993,HaydenWinter2003, anshu2022entanglement}.

The separation above distinguishes standardized entanglement from the nonuniform spectral structure that must remain in a quantum register. For given $b$ and $k$, the optimal reconstruction fidelity is \(F_{b,k}(\psi):=\max_{\hat U_A,\hat U_B,\gamma_k}\big|\langle\Omega_{b,k}(\gamma)| \hat U_A\otimes \hat U_B|\psi\rangle\big|^2.\) The optimization can be performed explicitly~\cite{Vidal2000,george2024reexamination}, giving the exact form
\begin{equation}
 F_{b,k}(\psi)=2^{-b}\sum_{j=0}^{2^k-1}
 \left(\sum_{s=0}^{2^b-1}\sqrt{\lambda_{j
 2^b+s}}\right)^2.
 \label{eq:frontier}
\end{equation}
See Supplemental Material (SM) for further details. For a prescribed $k$, the maximal fidelity over the Bell yields is $F_k:=\max_{0\le b\le n-k}F_{b,k}$. For a given error tolerance $\epsilon$, we define the minimum memory cost as 
\begin{equation}
    k_\epsilon:=\min\{k:F_k\ge1-\epsilon\}. 
    \label{eq:min_mem}
\end{equation}
At this minimum, the largest Bell yield compatible with the target fidelity is $b_\epsilon:=\max\{0\le b\le n-k_\epsilon:F_{b,k_\epsilon}\ge1-\epsilon\}$. The pair $(b_\epsilon,k_\epsilon)$ therefore describes a reconstruction that first minimizes residual memory, and then maximizes Bell yield while maintaining the desired error. Importantly, the compressibility problem considered here mainly concerns this minimum residual size \(k_\epsilon\), and not the computational cost of finding or implementing the encoders.

\textbf{\emph{Nonlocal stabilizer fidelity.}}
When no residual is allowed in Eq.~\eqref{eq:frontier}, \(k\!=\!0\), the target contains only flat Bell pairs and local pure factors. The optimized compression fidelity then reduces to the known nonlocal stabilizer fidelity~\cite{Sierant2026,Viscardi2026}, \(\FNL\!=\!F_{k=0}\!=\!\max_{0\le b\le n}2^{-b} \left(\sum_{i=0}^{2^b-1}\sqrt{\lambda_i}\right)^2\), where \(D\!:=\!-\log_2\FNL\) is \textit{nonlocal magic}. Allowing a residual, one obtains \(\FNL\le F_k\le\min\{1,(k+1)\FNL\}\). The upper fidelity bound follows by grouping ordered Schmidt amplitudes into successive powers of two, as described in SM. Crucially, if \(F_{\mathrm{NL}}\) tends to zero in the thermodynamic limit, then any memory size growing more slowly than \(1/F_{\mathrm{NL}}\) has vanishing fidelity. Further, requiring $F_k\ge1-\epsilon$, it holds that
\begin{equation}
 \begin{aligned}
 k_\epsilon&\ge\max\!\left\{0,\left\lceil(1-\epsilon)2^{D}-1\right\rceil\right\}.
 \end{aligned}
 \label{eq:memory_nlmagic_bound}
\end{equation}
Importantly, it follows that if \(D(n)-\log_2[k(n)+1]\to+\infty\), then \(F_{k(n)}\to0\). In particular, if \(D(n)=\beta\log_2 n+O(1)\) with \(\beta>0\), every memory size satisfying \(k(n)/n^\beta\to0\) yields vanishing fidelity. Maintaining fidelity at least \(f_0>0\) therefore requires \(k(n)\ge f_0 2^{D(n)}-1\), so the residual memory must grow at least proportionally to \(n^\beta\) for this example. Hence, unbounded growth of nonlocal magic constrains not only the best zero-memory approximation, but also the memory required to reach a prescribed reconstruction fidelity. Next, we show that smoothed entanglement spread directly bounds the residual memory required at a prescribed reconstruction accuracy.

\textbf{\emph{Entanglement spread.}} A complementary and more informative lower bound on the residual memory follows from \textit{entanglement spread}~\cite{HaydenWinter2003,Harrow2010,anshu2022entanglement}, which also constrains the communication required for entanglement dilution and pure-state preparation from Bell pairs. For $\hat\rho_A=\operatorname{Tr}_B|\psi\rangle\langle\psi|$, its unsmoothed form is
\begin{equation}
    \Delta_0(\psi):=\log_2\!\left(\operatorname{rank}\hat\rho_A\,\|\hat\rho_A\|_\infty\right),
    \label{eq:spread}
\end{equation}
where \(\|\hat\rho_A\|_\infty\) denotes the largest singular value. We can recast it as \(\Delta_0\!=\!S_0-S_\infty\), where \(S_\alpha\!:=\!(1-\alpha)^{-1}\log_2\operatorname{Tr}\hat\rho_A^\alpha\) is the \(\alpha\)-th R\'enyi entropy. \(\Delta_0\) is a measure of nonflatness of the reduced spectrum~\cite{karjula2026ebbs,PhysRevA.109.L040401}, and is additive under tensor products, so that tensoring flat Bell pairs leaves it unchanged. For a pure stabilizer state the nonzero Schmidt probabilities across every bipartition are equal, and one finds $S_0=S_\infty$; an exact factorization of the form Eq.~\eqref{eq:target} then satisfies \(\Delta_0(\psi)\!=\!\Delta_0(\gamma)\leq k_0\), meaning all the spread simply lives in the residual state (here \(k_0\equiv k_{\epsilon=0}\)). Crucially, the monotonicity of R\'enyi entropies also gives \(0\leq S_\alpha-S_\beta\leq\Delta_0\) for \(0\leq\alpha\leq\beta\leq\infty\), upper bounding all orders of R\'enyi entropy gaps. For power-of-two ranks it also (loosely) holds that \({D\le S_0-S_{1/2}\le S_0-S_\infty=\Delta_0}\), establishing that bounded spread prevents divergent nonlocal magic. For a full-rank state \(\Delta_0=n+\log_2\lambda_{\max}\).

The raw quantity \(\Delta_0\) is sensitive to arbitrarily small Schmidt weights and must be "smoothed" to describe a finite-error compression protocol. For the ordered Schmidt probabilities $\lambda$, we can define
\begin{equation}
    \Delta_\epsilon(\lambda):=
    \min_{\substack{\mu_r\\\sum_r\sqrt{\lambda_r\mu_r}\geq\sqrt{1-\epsilon}}}
    \log_2\!\left(|\operatorname{supp}\mu|\,\mu_{\max}\right),
    \label{eq:fidelity_smoothed_spread}
\end{equation}
where $\sum_r\mu_r=1$, and \(\mu\) ranges over normalized, nonincreasing probability vectors padded to dimension $2^n$. \(\operatorname{supp}\mu\) denotes the set of indices with positive entries, and \(\mu_{\rm max}\) is the largest entry. To remain consistent, we use the same squared fidelity convention throughout. Since every admissible factorized target has spread at most $k$, the minimum residual memory simply obeys
\begin{equation}
    \left\lceil\Delta_\epsilon(\lambda)\right\rceil
    \leq k_\epsilon(\psi).
    \label{eq:spread_memory_bounds}
\end{equation}
Physically, the above relation makes explicit that the residual memory \(\ket{\gamma_k}\) must accommodate the spectral nonflatness that cannot be discarded within the error tolerance. Overall, the spread bound, Eq.~\eqref{eq:spread_memory_bounds}, can give a stronger quantitative obstruction at a chosen accuracy, while the nonlocal magic bound, Eq.~\eqref{eq:memory_nlmagic_bound}, directly relates a resource measure to the full fidelity--memory tradeoff.

Given such characterizations, we next study how the spectral compressibility of dynamically generated states evolves in time in two distinct families of random quantum circuits.

\textbf{\emph{Chaotic random circuits}.} We first consider chaotic random unitary dynamics without conservation laws by evolving the initial state $\ket0^{\otimes N}$ on a periodic chain divided into equal contiguous halves. A time step consists of independent single-qubit Haar-random rotations followed by even-bond controlled-$\hat Z$ gates, \(\text {diag}(1,1,1,-1)\), then fresh random rotations followed by odd-bond entangling gates~\cite{Nahum2017,fisher2023random,PhysRevLett.123.190602}.

\begin{figure}[t]
\centering
\includegraphics[width=\columnwidth]{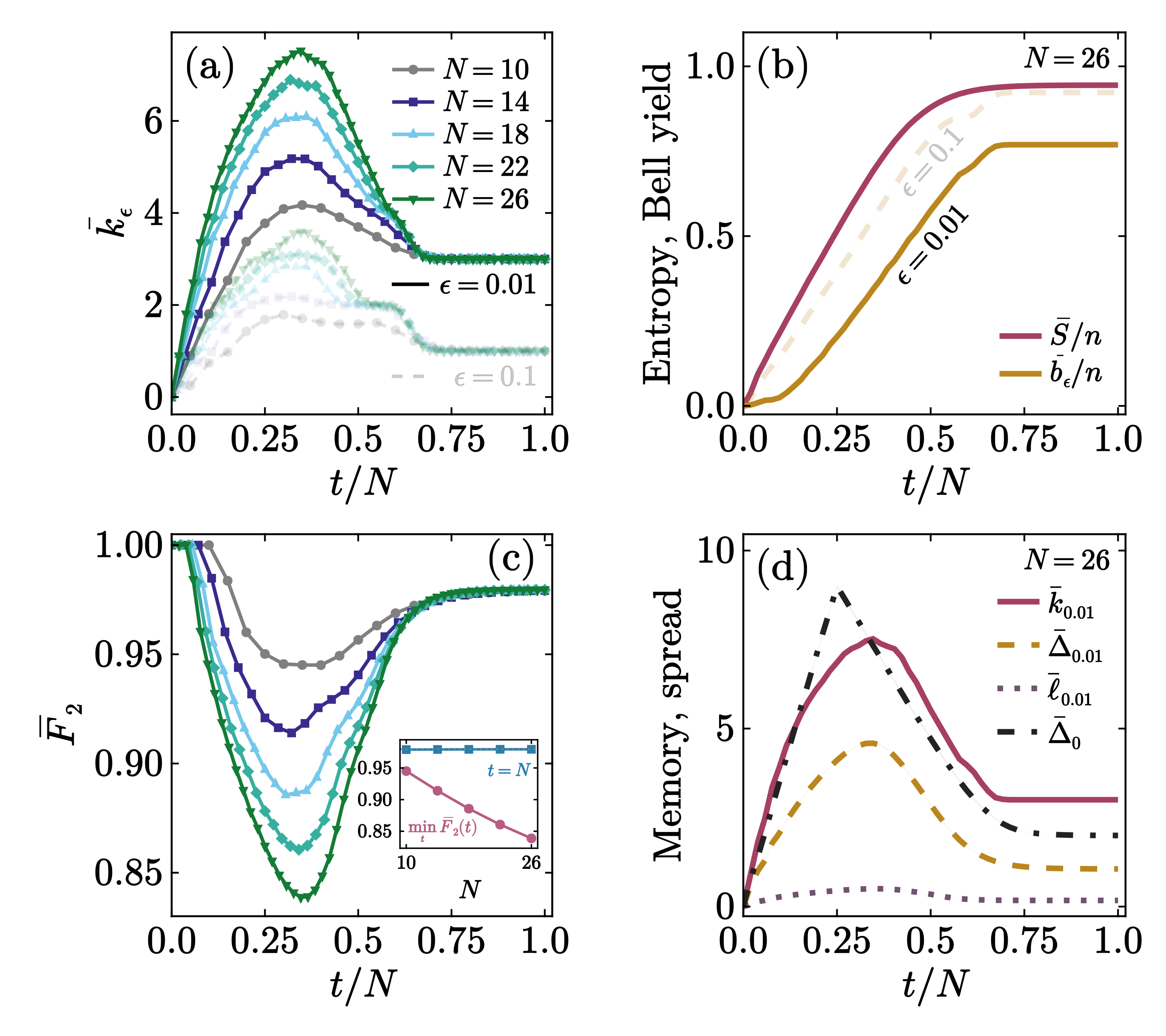}
\caption{\textbf{Spectral compression barrier in chaotic dynamics.}
\textbf{(a)} Mean minimum residual memory \(\bar k_{\epsilon}\) at $\epsilon=0.01$ (solid) and $0.10$ (faint dashed). \textbf{(b)} Average entropy density \(\bar S/n\), and finite-error Bell yield density \(\bar b_\epsilon/n\) at $N=26$, with the same error convention. \textbf{(c)} Optimal fixed-$k=2$ fidelity. Inset: its minimum ensemble mean at $t=\operatorname*{arg\,min}_{\tau}\overline{F}_2(\tau)$ (rose circles) and value at $t=N$ (blue squares), versus $N$. \textbf{(d)} \(\bar{k}_{\epsilon}\), mean smoothed spread \(\bar{\Delta}_{\epsilon}\), and mean magic bound $\bar{\ell}_\epsilon:=\max\{0,(1-\epsilon)\overline{2^D}-1\}$ at $N=26$, $\epsilon=0.01$, and mean raw spread $\bar{\Delta}_0$. Optimization precedes averaging, which is done over \(50\)--\(400\) independent circuit realizations. Error bars are of the size of the symbols or smaller.}
\label{fig:chaotic}
\end{figure}

As displayed in Figure~\ref{fig:chaotic}(a), the residual memory cost \(k_\epsilon\) first grows, reaches a size-dependent peak, and then decreases toward a common late-time plateau. Tightening the error tolerance \(\epsilon\) from $0.10$ to $0.01$ naturally raises the memory requirement while preserving this barrier-like pattern. During the later part of the evolution, the entropy continues to increase even as the remainder shrinks (Fig.~\ref{fig:chaotic}(b)). Entanglement can therefore continue to grow through the flat Bell-pair component even as the required nonflat residual memory decreases. Choosing a fixed register size \(k=2\), as shown in Fig.~\ref{fig:chaotic}(c) and its inset, the transient fidelity minimum decreases with $N$, whereas the late-time values are size independent and approach the high-fidelity Haar benchmark. These results give a compression interpretation of transient embezzling spectra associated with diverging nonlocal magic (\(D\to\infty\)) and entanglement capacity~\cite{Karjula2026,Aditya2026,karjula2026ebbs}, casting the ``magic barrier''~\cite{minar2026magic,zhang2026revealing} as an operational barrier to entanglement-spectrum compression.

In Figure~\ref{fig:chaotic}(d) we also plot the required memory with the lower bounds introduced earlier. At a fixed error, $\epsilon=0.01$, both the smoothed spread and nonlocal magic bound follow the transient rise and fall, while spread provides a tighter lower bound for these data. The numerical optimization of \(\Delta_\epsilon\) is described in SM. The behavior of \(\Delta_0\) also admits a simple interpretation. $S_0$ grows linearly until the rank saturates, while $S_\infty=-\log_2\lambda_{\max}$ continues to increase on average, reducing their difference and resulting in the rise-peak-fall behavior. For the chaotic transient embezzling states, the Schmidt spectra are already full-rank and nondegenerate, giving \(k_0=n\) and \(k_0/\Delta_0 = n(n-S_\infty)^{-1}\). Numerically, \(\overline{k_0/\Delta_0}\approx1.7\) over the sampled sizes, so the raw spread tracks a roughly fixed fraction of the exact residual memory.

\begin{figure}[t]
\centering
\includegraphics[width=\columnwidth]{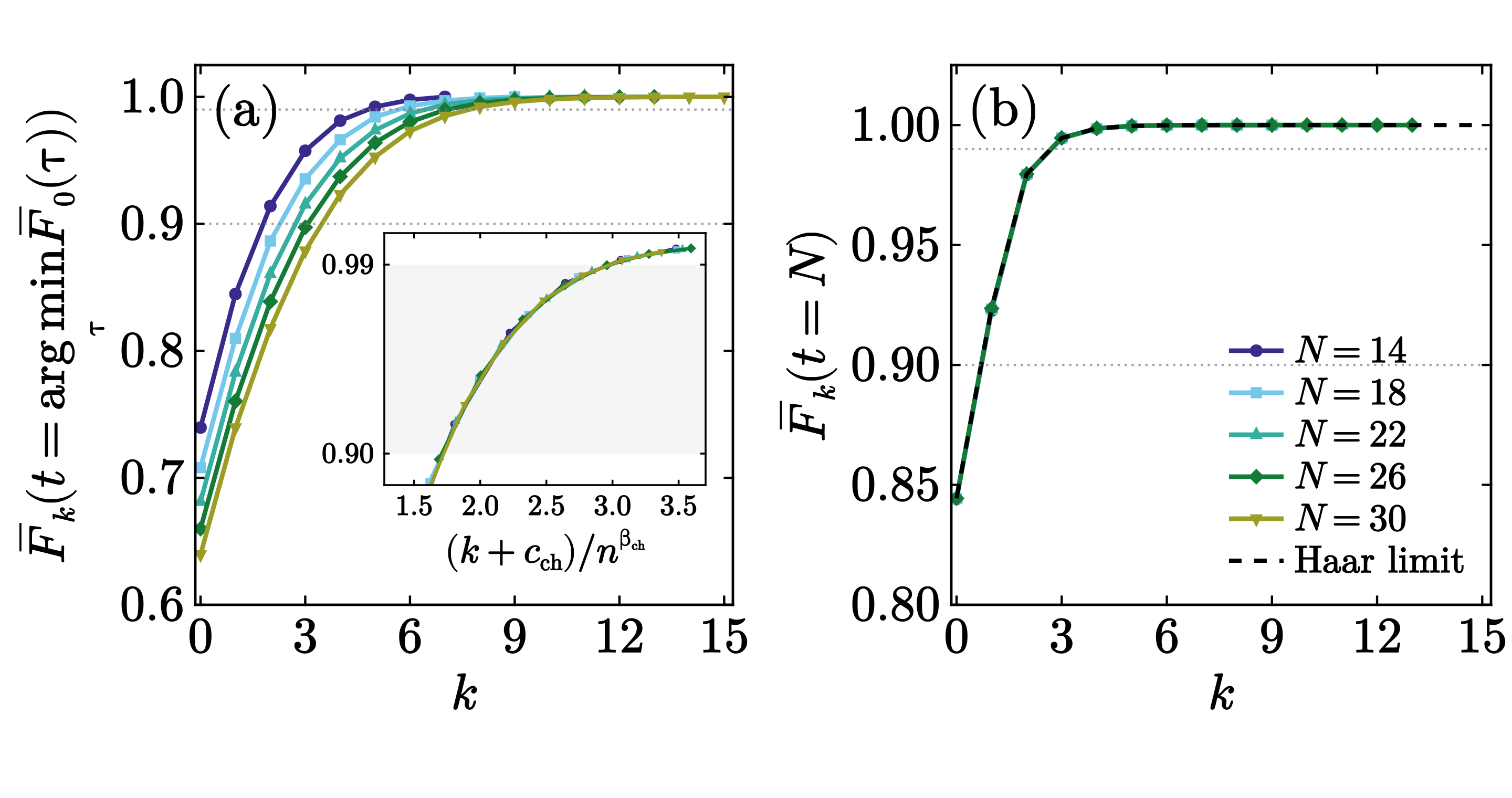}
\caption{\textbf{Compression fidelity--memory tradeoff in chaotic circuits.}
\textbf{(a)} Chaotic transient states evaluated at $t=\operatorname*{arg\,min}_{\tau}\overline{F}_0(\tau)$ for each size. Inset: collapse against $(k+c_{\rm ch})n^{-\beta_{\rm ch}}$, with $c_{\rm ch}\approx2.30$ and $\beta_{\rm ch}\approx0.45$, where "ch" denotes chaotic. \textbf{(b)} Late-time chaotic-circuit states at $t=N$. Black dashed line shows the analytically derived large-$N$ Haar limit. Each point is the ensemble mean of the individually optimized $F_k$ at integer $k$. Collapse parameters are fitted over the shaded window $0.90\leq\overline{F}_k\leq0.99$, denoted by the dotted horizontal lines. Data are averages over \(50\)--\(400\) independent circuit realizations and error bars are of the size of the symbols or smaller.}
\label{fig:frontiers}
\end{figure}

In Figure~\ref{fig:frontiers} we resolve the full fidelity--memory tradeoff for the chaotic transient and at the late time $t=N$ Haar states. The transient embezzling states approximately align under $(k+c_{\rm ch})n^{-\beta_{\rm ch}}$ with $\beta_{\rm ch}\approx0.45$ and $c_{\rm ch}\approx2.30$ (inset of Fig.~\ref{fig:frontiers}(a)) giving an effective and approximate finite-size exponent. The exponent \(\beta\) characterizes the growth of the memory required at fixed fidelity, while the size-independent offset \(c\) accounts phenomenologically for finite-size corrections. The approximate collapse is consistent with algebraic but subextensive growth of the residual memory over the sampled subsystem sizes in the high-fidelity regime. Remarkably, as hinted earlier, the fidelity–memory curves for Haar-random states instead coincide without rescaling the $k$ axis, approaching a size-independent form shown by the dashed line in Fig.~\ref{fig:frontiers}(b). This limiting curve follows from the Marchenko--Pastur spectral distribution~\cite{MarchenkoPastur1967,ZyczkowskiSommers2001,DumitriuEdelman2002} (see SM). Across a balanced cut, a typical large Haar state admits coherent compression into $b=n-O_\epsilon(1)$ Bell pairs with a residual register whose size depends only on the target accuracy. In this limit, the infidelity obeys $1-F_k^{\rm H}\leq\tfrac12\,4^{-k}$ for fixed $k\geq1$, so $O(\log(1/\epsilon))$ residual qubits per party suffice (see SM). With no residual memory, the optimum occurs at $b=n-1$, yielding $\FNL^{\rm H}\approx0.84$ and $D^{\rm H}\approx0.24$ bits. The raw spread also stays bounded, with $\Delta_0^{\rm H}=\log_2(2^n\lambda_{\max})\to2$~\cite{nechita2007asymptotics}. The nonzero limiting $\FNL^{\rm H}$ thus rules out universal embezzlement despite the near-maximal Bell yield~\cite{Sierant2026,Karjula2026}.

These characterizations also motivate reexamining the accuracy of the entropy-based estimate $\chi\sim2^{S_1}$ for an approximate matrix product state representation~\cite{PhysRevLett.93.040502,PhysRevLett.100.030504,Cao2025Gravity,Bao2019,Shaw2024}. Let $\chi_\epsilon$ be the smallest Schmidt rank that retains total probability at least $1-\epsilon$. We define the \textit{spectral compression gap} $\delta_\epsilon:=\log_2\chi_\epsilon-S_1$ to quantify the departure of the required bond dimension across the cut from this estimate. When \(\delta_\epsilon>0\), the Schmidt rank required at the chosen fidelity exceeds the entropy estimate \(2^{S_1}\) by a factor \(2^{\delta_\epsilon}\). To connect this gap to residual memory, note that a reconstruction with $b$ Bell pairs and $k$ residual qubits per side has Schmidt rank at most $2^{b+k}$. Achieving fidelity at least $1-\epsilon$ therefore requires $b+k\geq\lceil\log_2\chi_\epsilon\rceil=\lceil S_1+\delta_\epsilon\rceil$. In particular, the optimal allocation $(b_\epsilon,k_\epsilon)$ obeys $k_\epsilon\geq\max\{0,\lceil S_1+\delta_\epsilon\rceil-b_\epsilon\}$. This gives a necessary rank constraint on the optimal allocation. For exact reconstruction, $\epsilon=0$, each Bell pair contributes one bit to both $\log_2\chi_0$ and $S_1$. The gap therefore resides entirely in the residual state, $\delta_0(\psi)=\delta_0(\gamma)=S_0-S_1\leq\Delta_0$~\footnote{For power-of-two Schmidt ranks, one also has $D\leq S_0-S_{1/2}\leq\delta_0\leq\Delta_0$.}. If $\gamma$ has full Schmidt rank $2^k$, then $\delta_0=k-S_1(\gamma)$. As shown numerically in the SM, at fixed $\epsilon$, the required bond dimension for chaotic transient states exceeds the entropy-based estimate by a factor $2^{\delta_\epsilon}$ that increases over the sizes studied. At late times, this factor is approximately size-independent. These results reveal a growing multiplicative overhead associated with multiscale entanglement spectra that entropy alone misses.

\textbf{{\emph{Matchgate circuits.}}} We now consider free-fermionic dynamics, where independent nearest-neighbor Gaussian gates act in even and odd layers on a periodic ring, starting from the even-parity vacuum. Such matchgate dynamics admits an efficient (exactly solvable) covariance-matrix description~\cite{JozsaMiyake2008,PhysRevA.65.032325}. Each gate is a random $\mathrm{SO}(4)$ rotation on the four Majorana modes of its two sites, and the circuit preserves fermion parity but not particle number. Let $\Gamma_A$ denote the $2n\times 2n$ covariance matrix of subsystem $A$, defined by $(\Gamma_A)_{rs}:=(i/2)\langle[\hat c_r,\hat c_s]\rangle$, where $\hat c_r$ are its Majorana operators, normalized by $\{\hat c_r,\hat c_s\}=2\delta_{rs}$. For a pure bipartite fermionic Gaussian state, the eigenvalues of $\Gamma_A$ occur in pairs $\pm i\xi_j$, with $0\leq\xi_j\leq1$. The reduced state of a pure fermionic Gaussian state is unitarily equivalent to a product of single-mode density matrices, \(\hat\rho_A \simeq\bigotimes_{j=1}^{n}\operatorname{diag}\!\left((1+\xi_j)/{2},(1-\xi_j)/{2}\right)\)~\cite{peschel2003calculation,botero2004bcs, surace2021fermionic,collura2026nonlocal,iannotti2026non,magni2026entanglement}. Each $\xi_j$ describes one entangled mode pair and $0<\xi_j<1$ corresponds to a partially entangled pair.

In the invariant Gaussian ensemble, $\xi_j<1$ almost surely, so that $\operatorname{rank}\hat\rho_A=2^n$ and $\|\hat\rho_A\|_\infty=\prod_{j=1}^{n}(1+\xi_j)/2$, giving \(\Delta_0 =\sum_{j=1}^{n}\log_2(1+\xi_j)\). For an equal bipartition, the normalized single-mode spectral density of the invariant Gaussian ensemble is \(\rho_n(x)=(1/n)\sum_{\ell=0}^{n-1}(4\ell+1)P_{2\ell}(x)^2, 0\leq x\leq1\)~\cite{Bianchi2021}, where $P_m$ denotes the Legendre polynomial of degree $m$ and $\int_0^1\rho_n(x)\,dx=1$. Using this gives the exact finite-size stationary average, \(    \mathbb{E}_{\mathrm{ss}}\Delta_0=\sum_{\ell=0}^{n-1}(4\ell+1)\int_0^1 P_{2\ell}(x)^2\log_2(1+x)\,dx\). Note the steady state \(\mathbb{E}_{\mathrm{ss}}\Delta_0\) is evaluated for each pure state and then averaged over the ensemble. In the thermodynamic limit, the mode density approaches the arcsine density $\rho_\infty(x)=2/[\pi\sqrt{1-x^2}]$. Consequently,
\begin{align}
    \lim_{n\to\infty}
    \frac{\mathbb{E}_{\mathrm{ss}}\Delta_0}{n}
    &=
    \frac{2}{\pi\ln2}
    \int_0^1\frac{\ln(1+x)}{\sqrt{1-x^2}}\,dx
    \nonumber\\
    &=\frac{4G}{\pi\ln2}-1
      =0.682\ldots,
    \label{eq:gaussian_spread_density}
\end{align}
where $G=\sum_{m=0}^{\infty}(-1)^m/(2m+1)^2$ is Catalan's constant~\cite{Bianchi2021}. Thus the stationary unsmoothed spread has the leading behavior $\mathbb{E}_{\mathrm{ss}}\Delta_0\approx0.341\,N$. Such states have a nondegenerate Schmidt spectrum almost surely, so $k_0=n$ and \(k_0/\mathbb{E}_{\mathrm{ss}}\Delta_0\approx1.46\). 

\begin{figure}[t]
\centering
\includegraphics[width=\columnwidth]{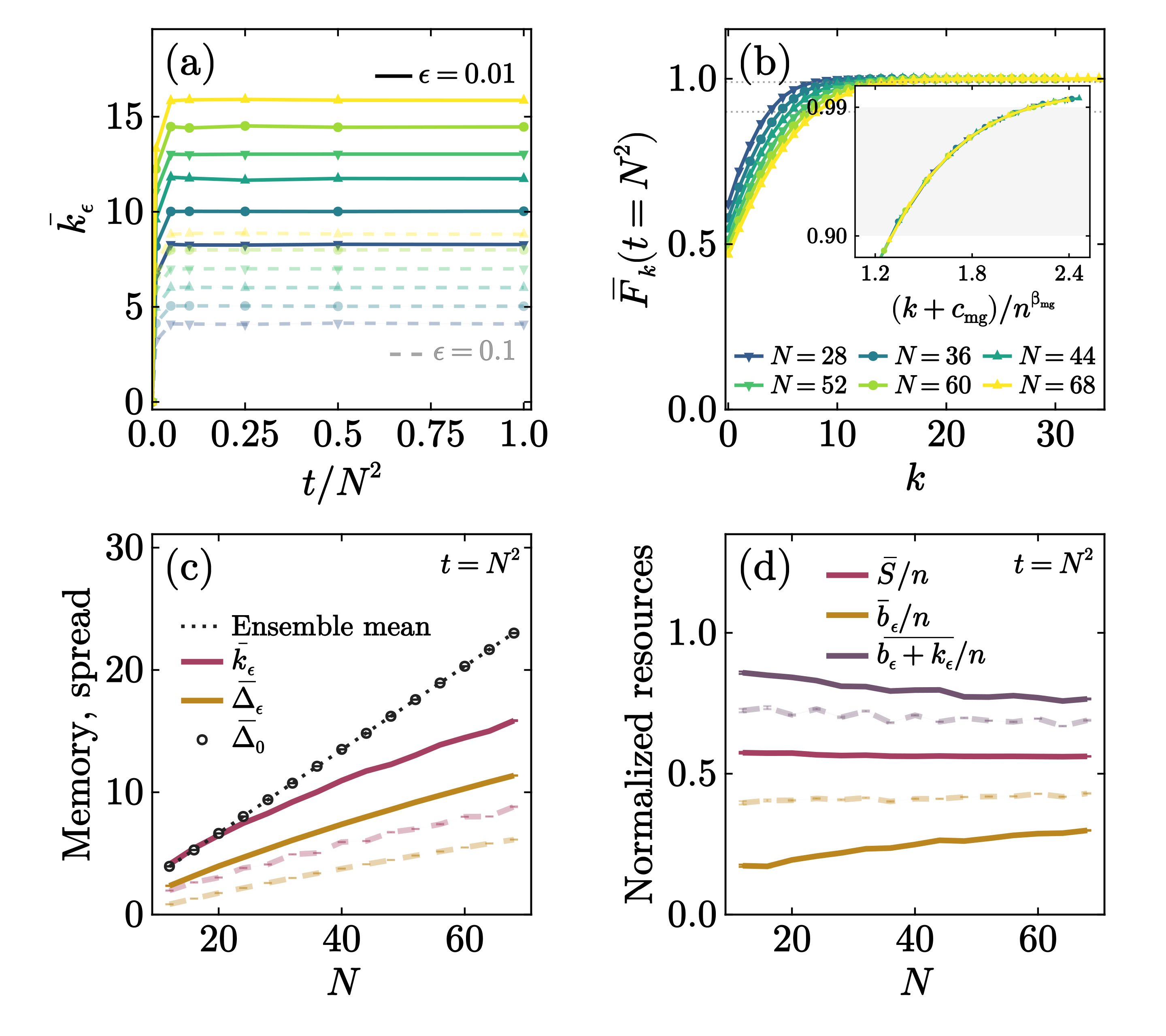}
\caption{\textbf{Persistent residual memory of stationary Gaussian states.}
\textbf{(a)} Mean minimum residual memory \(\bar k_{\epsilon}\) at $\epsilon=0.01$ (solid) and $0.10$ (faint dashed) as a function of $t/N^2$. \textbf{(b)} Stationary fidelity--memory frontiers (i.e. optimized over Bell-pair yield for each $k$). Inset: the same data against $(k+c_{\rm mg})n^{-\beta_{\rm mg}}$, with $c_{\rm mg}\approx2.48$ and $\beta_{\rm mg}\approx0.59$, where "mg" denotes matchgate. Collapse parameters are fitted over the shaded window
$0.90\leq\overline{F}_k\leq0.99$, denoted by the dotted horizontal lines. \textbf{(c)} Minimum memory and fidelity-smoothed spread at both errors, with the line convention of (a). Open circles show raw spread and the dotted curve is its exact stationary mean, Eq.~\eqref{eq:gaussian_spread_density}. \textbf{(d)} Bell yield, entropy, and total retained qubits per party, divided by $n=N/2$, with the same error convention. All optimizations precede averaging. Data are averages over \(100\)--\(200\) independent circuit realizations and error bars are of the size of the symbols or smaller.}
\label{fig:gaussian}
\end{figure}
As shown in Fig.~\ref{fig:gaussian}(a), here the late-time memory remains size dependent for the chosen error tolerances, which can be also verified by independently sampling the global $\mathrm{SO}(2N)$ rotations of the vacuum. The stationary fidelity--memory curves shift toward larger residual registers as the system grows (Fig.~\ref{fig:gaussian}(b)). Similar to the transient chaotic embezzlers, the high-fidelity parts roughly align against $(k+c_{\rm mg})n^{-\beta_{\rm mg}}$, with the numerical estimates $\beta_{\rm mg}\approx0.59$ and $c_{\rm mg}\approx 2.48$, fitted jointly across the displayed sizes. Noting that the stationary matchgate states are expected to obey the asymptotic scaling \(D=(1/2)\log_2n+O(1)\)~\cite{Aditya2026}, the fidelity bound Eq.~\eqref{eq:memory_nlmagic_bound} then implies that the residual memory must grow at least proportionally to $\sqrt n$ to retain a fixed positive fidelity. The fitted exponent \(\beta_{\rm mg}\) is thus consistent with the condition imposed by the growth of nonlocal magic (within the limits of the system sizes studied). Further, as shown in Fig.~\ref{fig:gaussian}(c), \(\Delta_{\epsilon}\) grows sublinearly alongside $k_\epsilon$ and bounds it from below; the linear growth of \(\Delta_0\) also agrees with the analytical average in Eq.~\eqref{eq:gaussian_spread_density}. Finally, Fig.~\ref{fig:gaussian}(d) shows the ensemble-averaged Bell-pair fraction $b_\epsilon/n$, total retained fraction $(b_\epsilon+k_\epsilon)/n$, and entropy density $S_1/n$. In the stationary Gaussian ensemble, both $b_\epsilon/n$ and $(b_\epsilon+k_\epsilon)/n$ are expected to approach the same asymptotic value as the entropy density $S_1/n\approx0.55$ in the thermodynamic limit \(N,n\to\infty\), consistent with extensive entanglement~\cite{Bianchi2021}.

\textbf{\emph{Conclusions.}}
The exact tradeoff between Bell-pair yield, residual memory, and reconstruction fidelity defines an operational measure of entanglement-spectrum complexity, $k_\epsilon$, quantifying the minimum nonflat residual memory required at a chosen accuracy. Smoothed entanglement spread bounds this cost from below, while diverging nonlocal magic forces it to grow without bound at any fixed fidelity. Thus, the spectral structure enabling universal entanglement embezzlement also obstructs faithful compression into Bell pairs and bounded residual memory. The residual memory cost in chaotic circuit dynamics rises to a peak and then decreases, even as entanglement entropy continues to grow. During this regime, the bond dimension required for fixed-fidelity compression across the cut exceeds the entropy based estimate by a factor whose peak increases over the simulated sizes. A large unrestricted Haar-typical state however asymptotically admits a size-independent remainder at fixed accuracy, and stationary fermionic Gaussian states retain a storage cost that increases with size. The optimal fidelity also provides a benchmark for quantum autoencoders~\cite{Romero2017} that separate Bell pairs from a residual register using local unitaries. Determining how efficiently such encoders may achieve this fidelity remains an open problem.

\textbf{\emph{Acknowledgments.}}
This work was supported by the European Union and the European Innovation Council through the Horizon Europe Project No. QRC-4-ESP (Grant Agreement No. 101129663), and EU Horizon Europe Quest project (Project No. 101156088). T.O. acknowledges the support by the Finnish Research Council project 362573 and the Finnish quantum flagship program. This work is part of the Finnish Center of Excellence in Quantum Materials (QMAT). GPT-6 family was used as an auxiliary tool for literature discovery, cross-checking scientific reasoning, and assisting with mathematical proofs, code development and automation. All scientific conclusions and interpretations are those of the authors.

\bibliography{ref}

\clearpage
\onecolumngrid

\begin{center}
    \textbf{\large Supplemental Material for}\\[0.5em]
    \textbf{"The magical memory cost of entanglement spectrum compression"}\\[0.5em]
    Matias Karjula, Vilma Nieminen, Teemu Ojanen, Tapio Ala-Nissila, and Moein N. Ivaki
\end{center}

\setcounter{equation}{0}
\setcounter{figure}{0}
\setcounter{table}{0}
\renewcommand{\theequation}{S\arabic{equation}}
\renewcommand{\thefigure}{S\arabic{figure}}
\renewcommand{\thetable}{S\arabic{table}}

\textbf{\emph{Optimal compression fidelity.}}
Let $b':=2^b$, and $k':=2^k$, and let $\mu_0\ge\cdots\ge\mu_{k'-1}$ be the residual Schmidt probabilities. Tensoring with $b$ Bell pairs repeats each $\mu_j/b'$ exactly $b'$ times. To maximize the overlap over local bases, we can write the two states \(\psi,\Omega\) in fixed product bases as coefficient matrices \(C_\psi\) and \(C_\Omega\). These are the matrices of amplitudes in fixed product bases, whose singular values are the Schmidt amplitudes. Local unitaries transform \(C_\psi\) into \(U_A C_\psi U_B^{\mathsf T}\), and the overlap becomes \(\operatorname{Tr}(C_\Omega^\dagger U_A C_\psi U_B^{\mathsf T})\), where \(U_{A,B}\) are the matrix representations of \(\hat U_{A,B}\). The von Neumann trace inequality bounds the absolute value of this trace by the sum of products of the ordered singular values. Equality can then be achieved by aligning the corresponding left and right singular vectors, which the independent local unitaries allow~\cite{Vidal2000}. Thus the optimal overlap pairs the largest Schmidt amplitude of one state with the largest of the other, and so on. Applying these results, for a fixed residual spectrum \(\mu\) it gives
\begin{equation}
 \max_{\hat U_A,\hat U_B}|\langle\Omega_{b,k}|(\hat U_A\otimes \hat U_B)|\psi\rangle|
 ={\frac{1}{\sqrt{b'}}}\sum_{j=0}^{k'-1}A_j\sqrt{\mu_j},
 \label{eq:overlap-proof}
\end{equation}
where $A_j:=\sum_{s=0}^{b'-1}\sqrt{\lambda_{jb'+s}}$. Using the Cauchy--Schwarz inequality then gives
\begin{equation}
 \left(\sum_j A_j\sqrt{\mu_j}\right)^2
 \le\left(\sum_jA_j^2\right)\left(\sum_j\mu_j\right)=\sum_jA_j^2.
 \label{eq:cs}
\end{equation}
Equality is achieved by taking $\sqrt{\mu_j}$ proportional to $A_j$, or \(\mu_j=(A_j^2)/\sum_l A_l^2\), where the denominator normalizes the residual. This choice also respects the ordering, as $A_j$ are nonincreasing and the input probabilities are sorted. Local unitary encoders that map the corresponding Schmidt bases to the designated register bases then achieve the equality. Thus Eq.~\eqref{eq:frontier} in the main text is an attainable optimum, and not only an upper bound.

It is also useful to note that the optimal reconstruction error separates exactly into a discarded Schmidt weight and a flattening error, where the errors arise from approximating the encoded state by the prescribed Bell-pair-plus-remainder form. We can write \(1-F_{b,k}=\delta_{\mathrm{tail}}+\delta_{\mathrm{flat}}\), where
\begin{align}
\delta_{\mathrm{tail}}&:=\sum_{r=b'k'}^{2^n-1}\lambda_r,\\\nonumber
\delta_{\mathrm{flat}}&:=\sum_{j=0}^{k'-1}\left[\sum_{s=0}^{b'-1}\lambda_{jb'+s}
-\frac{1}{b'}\left(\sum_{s=0}^{b'-1}\sqrt{\lambda_{jb'+s}}\right)^2\right].
\end{align}
Here $\delta_{\mathrm{tail}}$ is the probability weight beyond the retained Schmidt components. The second contribution measures the departure from the equal-weight blocks required by $b$ Bell pairs. Indeed, defining \(\bar a_j:=(1/{b'})\sum_{s=0}^{b'-1}\sqrt{\lambda_{jb'+s}},\) one obtains \(\delta_{\mathrm{flat}}=\sum_{j=0}^{k'-1}\sum_{s=0}^{b'-1}\left(\sqrt{\lambda_{jb'+s}}-\bar a_j\right)^2\ge0. \) Thus, as one might expect, flattening incurs no error precisely when the probabilities within each retained block are equal. For \(b=0\), each block contains a single probability and \(\delta_{\mathrm{flat}}=0\). For \(b+k=n\), all Schmidt components are retained and \(\delta_{\mathrm{tail}}=0\), although flattening can still incur an error when \(b>0\).

\textbf{\emph{Memory bound from fidelity.}}
Adopting the same notations, $T_j:=A_j/\sqrt{b'}$ gives a nonnegative, nonincreasing sequence with $F_{b,k}=\sum_{j=0}^{k'-1}T_j^2$. The first term, $T_0^2$, is the fidelity with a rank-$b'$ Bell sector and is therefore at most $\FNL$. The remaining terms can be bounded in groups with indices $m,\ldots,2m-1$, where $m=1,2,4,\ldots,2^{k-1}$. Since the sequence is nonincreasing, every $T_j$ in such a group is no larger than the average of the first $m$ terms. Consequently,
\begin{align}
 \sum_{j=m}^{2m-1}T_j^2
 &\le m\left(\frac1m\sum_{j=0}^{m-1}T_j\right)^2\nonumber\\
 &=\frac1{b'm}\left(\sum_{i=0}^{b'm-1}\sqrt{\lambda_i}\right)^2
 \le\FNL.
 \label{eq:shell-proof}
\end{align}
Here the first $m$ blocks contain exactly the first $b'm$ Schmidt amplitudes, and the resulting expression is therefore a Bell-sector fidelity of rank $b'm$. The choice of power-of-two group sizes ensures that $b'm=2^{b+\log_2m}$ is an allowed Bell-sector rank. The first term $T_0^2$ plus these $k$ groups cover all $2^k$ terms, with each contributing at most $\FNL$. Summing gives $F_{b,k}\le(k+1)\FNL$, and maximizing over $b$ proves the upper bound on $F_k$. For the lower bound, we simply note that any zero-memory approximation can also be represented with a $k$-qubit residual by moving Bell pairs into the residual register, and filling its remaining qubits with local pure factors. Thus $F_k\ge\FNL$. Finally, requiring $F_k\ge1-\epsilon$ gives $1-\epsilon\le(k+1)\FNL$. Rearranging and noting that $k$ is an integer, yields the finite-error memory bound in Eq.~\eqref{eq:memory_nlmagic_bound}.

\textbf{\emph{Compression of unrestricted Haar-random states.}}
The key idea here is to eventually express the Haar compression problem as an approximation of a function by its averages over equal intervals. For $d=2^n$, the rescaled Schmidt probabilities $x:=d\lambda$ have density $\rho(x):=(2\pi)^{-1}\sqrt{(4-x)/x}$ on $0<x<4$ in the balanced (equal contiguous bipartitioning) Haar-typical limit~\cite{MarchenkoPastur1967,ZyczkowskiSommers2001}. We order the probabilities from largest to smallest, and let $u\in[0,1]$ denote the fraction counted from the largest; $x(u)$ then gives the rescaled probability at position $u$ in the ordered spectrum, with $u=0$ corresponding to the largest values and $u=1$ to the end of the spectrum. Equivalently, $u=\int_{x(u)}^4\rho(y)\,dy$. The substitution $x=4\cos^2\theta$ can simplify the square root in the density, giving
\begin{equation}
 x(u)=4\cos^2\theta,\quad
 u=\frac{2\theta-\sin2\theta}{\pi},
 \quad 0\leq\theta\leq\frac\pi2.
\end{equation}
Using $du/d\theta=4\sin^2\theta/\pi$, the cumulative
amplitude $X(u)=\int_0^u\sqrt{x(v)}\,dv$ and cumulative
probability $P(u)=\int_0^u x(v)\,dv$ become
\begin{equation}
 X(u)=\frac8{3\pi}\sin^3\theta,
 \quad
 P(u)=\frac{2\theta-\tfrac12\sin4\theta}{\pi}.
\end{equation}
Here $u$ counts eigenvalues, whereas $P(u)$ counts their probability weight. For example, the largest half of the Schmidt probabilities carries the majority of the total probability, and $P(1/2)\approx0.8937$.

For an allocation $(b,k)$, write $r:=n-b-k$ for the number of local pure-state qubits per party. The target has Schmidt rank at most $2^{b+k}$, so its optimal overlap uses the largest $2^{b+k}$ input probabilities. These occupy a fraction $2^{-r}$ of the full spectrum and are grouped into $2^k$ blocks, each containing $2^b$ probabilities. Each block therefore has width $h:=2^b/d=2^{-(k+r)}$ in the variable $u$. In the large-$d$ limit, its amplitude sum becomes $\sum_{\mathrm{block}}\sqrt{\lambda_i}
\simeq\sqrt d\int_{\mathrm{block}}\sqrt{x(u)}\,du$. Substituting into Eq.~\eqref{eq:frontier}, with $d/2^b=1/h$, gives the limiting fidelity at fixed $k,r$,
\begin{equation}
 \mathcal F_{k,r}
 =\frac1h\sum_{j=0}^{2^k-1}
 [X((j+1)h)-X(jh)]^2.
 \label{eq:haar-allocation}
\end{equation}
This expression accounts for both the omitted probability beyond $u=2^{-r}$ and the error from approximating the retained amplitudes by equal values within each block. In particular, the fidelity cannot exceed the probability included in the target support, so $\mathcal F_{k,r}\leq P(2^{-r})$.

We may now try and optimize the Bell yield explicitly by comparing the numerical values of fidelity for allowed choices of $r$. The numerical values can be calculated directly using Python/SciPy or similar libraries. With no residual memory \(k=0\), choosing $r=0$ and $r=1$ give $\mathcal F_{0,0}\approx0.7205$ and $\mathcal F_{0,1}\approx0.8444$, respectively. Every $r\geq2$ includes at most the largest quarter of the spectrum and therefore has fidelity at most $P(1/4)\approx0.6229$. Thus $r=1$, or $b=n-1$, is optimal for $k=0$. With one residual qubit $k=1$, retaining the whole spectrum instead gives $\mathcal F_{1,0}\approx0.9236$. This already exceeds $P(1/2)\approx0.8937$, the upper bound for every allocation with $r\geq1$. Increasing $k$ at $r=0$ subdivides each block into smaller blocks, allowing a more accurate approximation of the amplitudes and hence a fidelity that cannot decrease. Therefore, $r=0$, or $b=n-k$, remains optimal for every fixed $k\geq1$. Allocations with $r$ growing without bound as $n$ increases cannot improve the result either. Their retained probability is at most $2^{-r}d\lambda_{\max}$, which vanishes since $d\lambda_{\max}\to4$~\cite{nechita2007asymptotics}. Setting $r=0$, and hence $h=2^{-k}$, in Eq.~\ref{eq:haar-allocation} then gives $F_k^{\rm H}=2^k\sum_{j=0}^{2^k-1} \big[X((j+1)2^{-k})-X(j2^{-k})\big]^2$.

For the optimal choice $r=0$, no Schmidt components are omitted, so the infidelity arises entirely from flattening the amplitudes within each block. To bound this error, let $f(u)=\sqrt{x(u)}$ and set $h=2^{-k}$. The mean amplitude on block $j$ is $\bar f_j=[X((j+1)h)-X(jh)]/h$, so $F_k^{\rm H}=\sum_j h\bar f_j^2$. Since $\int_0^1 f(u)^2\,du=1$, subtracting these squared block means gives
\begin{equation}
 1-F_k^{\rm H}
 =\sum_{j=0}^{2^k-1}\int_{jh}^{(j+1)h}
       [f(u)-\bar f_j]^2\,du.
\end{equation}
Thus the infidelity measures how much the amplitudes vary within the blocks. On each interval of width $h$, the Poincar\'e-Wirtinger inequality bounds this variation by $h^2/\pi^2$ times the integral of $|f'(u)|^2$. Summing over all blocks gives
\begin{equation}
 1-F_k^{\rm H}
 \leq\frac{h^2}{\pi^2}\int_0^1|f'(u)|^2\,du
 =\frac12\,4^{-k},\quad k\geq1.
 \label{eq:haar-error}
\end{equation}
To evaluate the integral, the parametrization above gives $f'(u)=-\pi/(2\sin\theta)$ and $du=4\sin^2\theta\,d\theta/\pi$. Consequently, $\int_0^1|f'(u)|^2\,du =\pi\int_0^{\pi/2}d\theta=\pi^2/2$. Although $f'(u)$ diverges at $u=0$, its square is integrable with respect to $u$, so the inequality applies. Each additional residual qubit halves the block width and reduces this error bound by a factor of four.

These statements concern the limiting Haar spectrum, with $n\to\infty$ taken at fixed $k$. For any fixed $\epsilon>0$, one can choose a fixed $k\geq1$ such that $\tfrac12 4^{-k}<\epsilon$. The reconstruction fidelity then exceeds $1-\epsilon$ with probability approaching one as $n$ increases. As promised, the limiting memory only depends on the chosen accuracy.

\begin{figure}[t]
 \centering
 \includegraphics[width=\columnwidth]{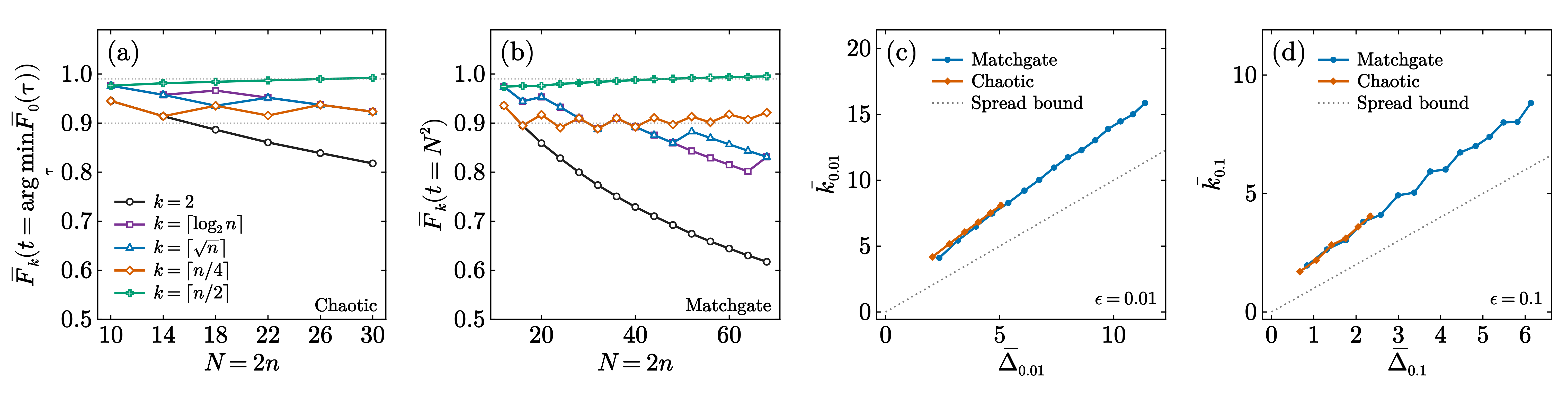}
 \caption{\textbf{Different memory budgets and the spread lower bound.}
 (a,b) Mean fidelity for $k=2$, $k=\lceil\log_2 n\rceil$, $k=\lceil\sqrt n\rceil$, $k=\lceil n/4\rceil$, and $k=\lceil n/2\rceil$, with $N=2n$. (a) Chaotic states at a common pilot-selected time $t= \arg\min_\tau \overline{F}_0 (\tau)$ for each size. (b) Independent stationary Gaussian states. Bell yield is optimized per state and budget. Dotted lines mark $0.90$ and $0.99$; error bars give pointwise $95\%$ state-bootstrap intervals, conditional on the selected time in (a). Open markers overlap when integer budgets coincide. (c,d) Mean minimum memory versus mean optimized smoothed spread at $\epsilon=0.01$ and $0.10$, respectively. Orange diamonds: chaotic transient; blue circles: stationary matchgate. Each point is one size, with both quantities computed per state before averaging. Dotted lines give the unrounded lower bound $\bar{k}_{\epsilon}\geq\bar{\Delta}_{\epsilon}$}
 \label{fig:budgets}
\end{figure}

\textbf{\emph{Additional results.}} Figure~\ref{fig:budgets} compares fixed and size-dependent residual-memory budgets in the two circuit models. For chaotic circuits, all budgets are evaluated at the same sampled time minimizing $\overline{F}_0$ for each system size. For matchgate circuits, the states are evaluated at $t=N^2$. Figures~\ref{fig:budgets}(a,b) show that the fidelity obtained with a fixed budget $k=2$ decreases with increasing system size. Allowing the residual register to grow improves reconstruction, but the different budget choices have distinct finite-size behavior. In particular, the logarithmic and square-root budgets also show decreasing fidelity over the larger matchgate sizes, whereas the extensive budgets retain higher fidelities over the sampled range. Steps and overlaps in these curves arise from rounding the budgets to integer numbers of qubits. Figures~\ref{fig:budgets}(c,d) compare the minimum residual memory directly with the optimized smoothed spread at the same error tolerance. The chaotic transient and stationary matchgate data follow similar curves over their overlapping range, while satisfying the spread lower bound $\overline{k}_\epsilon\geq\overline{\Delta}_\epsilon$.

We further examine the accuracy of entropy-based estimates of spectral compressibility, following the discussions in Ref.~\cite{Cao2025Gravity}. For the ordered Schmidt probabilities, let $\chi_\epsilon=\min\{\chi\,|\,\sum_{r=0}^{\chi-1}\lambda_r\geq1-\epsilon\}$ and define the \textit{spectral compression gap} as $\delta_\epsilon=\log_2\chi_\epsilon-S(\hat\rho_A)$. The retained weight is the squared fidelity of the optimal normalized Schmidt-truncated state, so $\epsilon$ uses the same reconstruction-error convention as the compression task in the main text. Since $\chi_\epsilon=2^{S+\delta_\epsilon}$, a positive gap quantifies the multiplicative excess $2^{\delta_\epsilon}$ of the required Schmidt rank over the entropy estimate $2^{S}$. Thus \(\delta_\epsilon\) measures the correction beyond entropy to the total retained Schmidt space, and therefore determines the minimum bond dimension required at this cut in an open-boundary matrix-product-state representation. Such an overhead can grow when a fixed fraction of Schmidt probability, exceeding the error tolerance $\epsilon$, lies at entanglement energies $-\log_2\lambda_r$ increasingly far above their mean $S_1$. Retaining this weight requires an increasing number of individually small Schmidt probabilities. 

\begin{figure}[t]
 \centering
 \includegraphics[width=\columnwidth]{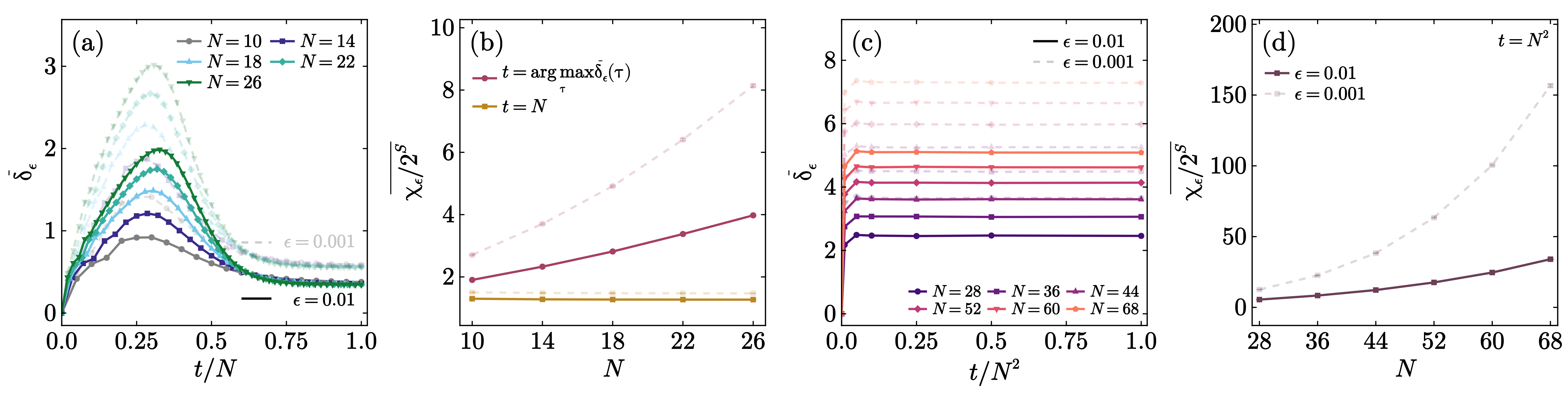}
 \caption{\textbf{Spectral compression gap in chaotic and matchgate dynamics.}
\textbf{(a)} Ensemble mean of $\delta_\epsilon=\log_2\chi_\epsilon-S$ during chaotic circuit evolution, where $\chi_\epsilon$ is the smallest Schmidt rank retaining total probability at least $1-\epsilon$ across the contiguous equal-half cut. \textbf{(b)} Mean of the statewise rank ratio $\chi_\epsilon/2^{S(\rho_A)}=2^{\delta_\epsilon}$ at the sampled maximum-gap time $t=\operatorname*{arg\,max}_{\tau}\overline{\delta}_\epsilon(\tau)$, selected separately for each system size and error tolerance, and at late time $t=N$. \textbf{(c)} Corresponding ensemble-mean gap during matchgate circuit evolution, plotted against $t/N^2$. \textbf{(d)} Mean statewise rank ratio for matchgate states at the measured late-time checkpoint $t=N^2$, as a function of $N$. Solid curves show $\epsilon=0.01$, and faint dashed curves show $\epsilon=0.001$ in all panels.}
\label{fig:spectral_gap}
\end{figure}

Figure~\ref{fig:spectral_gap} shows a transient peak in \(\delta_\epsilon\) that increases over the sampled sizes and recedes toward the size-independent Haar limits. We compute \(\chi_\epsilon\) from the state evolved on the periodic ring. In an open-boundary matrix product state ordered along the chain, \(\chi_\epsilon\) is the minimum dimension of the single bond crossing the half-chain bipartition at the prescribed fidelity, with all other bonds unrestricted. The corresponding ratio \(\chi_\epsilon/2^{S_1}=2^{\delta_\epsilon}\) also grows with size at the peak, suggesting a stretched-exponential overhead relative to the entropy-based estimate. Tightening the error tolerance from \(0.01\) to \(0.001\) increases the gap, as more of the small Schmidt weights must be retained. Overall, these results indicate that entropy becomes a progressively poorer predictor of the rank required at fixed accuracy around the transient regime, while its relative accuracy improves in the Haar-typical regime $t=N$. For matchgate circuits, the overhead remains size dependent at $t=N^2$.  Since these circuits remain efficiently simulable using fermionic covariance matrices, a large matrix-product-state representation cost does not by itself imply classical computational hardness.

\textbf{\emph{Numerical evaluation of the fidelity-smoothed spread.}} For each simulated pure state, our goal is to find the nearby Schmidt spectrum with the smallest spread. We denote the fixed Schmidt probabilities of the simulated state by $\lambda_i$, ordered as $\lambda_1\ge\cdots\ge\lambda_{D_\lambda}>0$, with $\sum_i\lambda_i=1$. The probabilities $\mu_i$ describe a trial
pure state and are varied in the optimization. When the Schmidt bases are aligned and both spectra are ordered, their squared fidelity is $F(\lambda,\mu)=(\sum_i\sqrt{\lambda_i\mu_i})^2$. For a prescribed minimum fidelity $f=1-\epsilon$, we compute
\begin{equation}
\Delta_\epsilon(\lambda)
=
\min_{\substack{\mu_i\ge0,\ \sum_i\mu_i=1\\
                 F(\lambda,\mu)\ge f}}
\log_2\!\left(|\operatorname{supp}\mu|\,\max_i\mu_i\right).
\label{eq:ss-objective}
\end{equation}
The two factors in the objective are the number of positive $\mu_i$ and the largest $\mu_i$. Retaining more components can reduce the largest $\mu_i$, so both factors must be optimized.

\paragraph{Fidelity-maximizing spectrum at a given rank and cap.}

Consider a candidate rank $R=|\operatorname{supp}\mu|$ and impose a common upper bound $c$ on the trial probabilities, $\mu_i\le c$. Normalization requires $c\ge1/R$. To determine whether this pair $(R,c)$ can attain the required fidelity, we find the largest fidelity achievable under these constraints. The $R$ positive $\mu_i$ can be aligned with the $R$ largest $\lambda_i$, since pairing a positive $\mu_i$ with a larger $\lambda_i$ increases its contribution to the overlap. Maximizing the fidelity is equivalent to maximizing the overlap amplitude $H(\mu)=\sum_{i=1}^{R}\sqrt{\lambda_i\mu_i}$. Consider two trial probabilities below the cap. A small amount of probability can be transferred between them in either direction while preserving normalization. At the maximum, their derivatives $\partial H/\partial\mu_i=\sqrt{\lambda_i}/(2\sqrt{\mu_i})$ must therefore be equal. This gives $\mu_i=\alpha\lambda_i$ with the same $\alpha$ for every uncapped $\mu_i$. For a capped component, $\mu_i=c$, probability can still be transferred from it to an uncapped component. Such a transfer would improve the overlap if the derivative at the capped component were smaller than the common derivative at the uncapped components. The latter derivative is $1/(2\sqrt{\alpha})$, so optimality requires $\sqrt{\lambda_i}/(2\sqrt c)\ge1/(2\sqrt{\alpha})$, or $\alpha\lambda_i\ge c$. Combining the two cases gives
\begin{equation}
\mu_i^{(R,c)}
=
\begin{cases}
\min\{c,\alpha\lambda_i\}, & i\le R,\\
0, & i>R,
\end{cases}
\qquad
\sum_{i=1}^{R}\min\{c,\alpha\lambda_i\}=1.
\label{eq:ss-capped-spectrum}
\end{equation}
The normalization equation determines how much the uncapped $\lambda_i$ must be scaled after truncation and capping. Concavity of $H$ makes these conditions sufficient for a global maximum. At $c=1/R$, normalization directly gives $\mu_i=1/R$ for all $i\le R$. We write $F_{\max}(R,c)=F(\lambda,\mu^{(R,c)})$ for the fidelity of the maximizing spectrum.

\paragraph{Computing the smoothed spread.}

We first determine which trial ranks $R$ can reach the fidelity threshold. Let $P_R=\sum_{i=1}^{R}\lambda_i$ be the mass of the $R$ largest original Schmidt probabilities. Cauchy--Schwarz gives $F(\lambda,\mu)\le P_R$ for every normalized $\mu$ on these indices, with equality for $\mu_i=\lambda_i/P_R$. The first feasible rank is therefore $R_{\min}=\min\{R:P_R\ge f\}$. For each feasible $R$, let $m_R$ denote the smallest cap for which $F_{\max}(R,m_R)\ge f$. Then
\begin{equation}
\Delta_\epsilon(\lambda)
=
\log_2\!\min_{R_{\min}\le R\le D_\lambda} Rm_R.
\label{eq:ss-rank-optimization}
\end{equation}
Every integer rank in this range is eligible. The rank that minimizes $Rm_R$ determines the optimal trial spectrum through Eq.~\eqref{eq:ss-capped-spectrum}. To compute $m_R$, we first test the uniform trial spectrum $\mu_i=1/R$ for $i\le R$. Its fidelity is $(\sum_{i=1}^{R}\sqrt{\lambda_i})^2/R$. If this reaches $f$, then $m_R=1/R$ and the spread is zero. Otherwise the smallest feasible cap satisfies $F_{\max}(R,m_R)=f$. We find this cap by identifying how many trial probabilities $\mu_i$ are capped when the fidelity reaches $f$. Denote this number by $j$. Since the $\lambda_i$ are ordered, the first $j$ trial probabilities equal $c$, while the remaining $R-j$ satisfy $\mu_i=\alpha\lambda_i$. Define $A_j=\sum_{i=1}^{j}\sqrt{\lambda_i}$ and $T_{j,R}=\sum_{i=j+1}^{R}\lambda_i$. The first $j$ trial probabilities carry mass $jc$, so normalization of $\mu$ gives $\alpha T_{j,R}=1-jc$. The two groups then contribute to the overlap as
\begin{equation}
\sqrt{F_{\max}(R,c)}
=
A_j\sqrt c+\sqrt{\alpha}\,T_{j,R}
=
A_j\sqrt c+\sqrt{T_{j,R}(1-jc)}.
\label{eq:ss-overlap-branch}
\end{equation}

To identify $j$, consider the boundary at which the $j$th trial probability first reaches the cap as $c$ is lowered. At this boundary, $c=\alpha\lambda_j$. Substituting into the normalization
condition gives the boundary cap $\eta_j^{(R)}$ and its overlap amplitude $g_j^{(R)}$:
\begin{equation}
\eta_j^{(R)}
=
\frac{\lambda_j}{j\lambda_j+T_{j,R}},
\qquad
g_j^{(R)}
=
\sqrt{\eta_j^{(R)}}
\left(A_j+\frac{T_{j,R}}{\sqrt{\lambda_j}}\right).
\label{eq:ss-breakpoints}
\end{equation}
Lowering the cap restricts the allowed $\mu_i$, so the boundary overlaps are nonincreasing. The first boundary has $g_1^{(R)}=\sqrt{P_R}$, while the last has $g_R^{(R)}=A_R/\sqrt R$, the uniform-spectrum overlap. After the uniform test, a binary search over the integer $j$ therefore finds $g_j^{(R)}\ge\sqrt f>g_{j+1}^{(R)}$. The desired cap lies between these two boundaries. With $j$ determined, we set Eq.~\eqref{eq:ss-overlap-branch} equal to $\sqrt f$ and solve directly for the cap. Writing $x=\sqrt c$ gives the quadratic equation
\begin{equation}
(A_j^2+jT_{j,R})x^2
-2A_j\sqrt f\,x+f-T_{j,R}=0.
\label{eq:ss-cap-quadratic}
\end{equation}
The smaller admissible root gives $m_R=x^2$ in the interval $[\eta_{j+1}^{(R)},\eta_j^{(R)}]$. Thus the integer search determines how many $\mu_i$ are capped, and the quadratic determines the cap itself. For chaotic states, the complete ordered $\lambda$ spectrum is available. We precompute $A_j$ and the suffix masses $Q_j=\sum_{i=j+1}^{D_\lambda}\lambda_i$, accumulating the latter from the smallest $\lambda_i$. Each boundary evaluation then uses $T_{j,R}=Q_j-Q_R$. We evaluate all feasible ranks in batches and take the minimum in Eq.~\eqref{eq:ss-rank-optimization}.

\paragraph{Additional optimizations for matchgate states.}

For matchgate states, we perform the same optimization over $\mu$. The Gaussian structure of the original state allows us to access its $\lambda_i$ without first constructing the full spectrum. After removing exactly pure modes, each of the $q$ remaining modes contributes two probabilities $p_a^\pm=(1\pm\xi_a)/2$. The $2^q$ Schmidt probabilities $\lambda_i$ are the products obtained by choosing one factor $p_a^+$ or $p_a^-$ from each mode.

We first use the mode factors to test whether a rank-one or full-support uniform trial spectrum $\mu$ already reaches $f$. Either case gives zero spread. Otherwise, we split the modes into two groups of approximately equal size and form the spectrum of each group. Each $\lambda_i$ is then a product of one entry from each list. By counting products above a threshold in the sorted lists, we select the $K$ largest $\lambda_i$. These entries retain their original probabilities, so $P_K=\sum_{i=1}^{K}\lambda_i$ is their mass in the full spectrum. Here $K$ counts the original probabilities $\lambda_i$ available for the calculation, whereas $R$ counts the positive trial probabilities $\mu_i$. If $P_K<f$, even rank $K$ cannot reach the target and we enlarge $K$. Once $P_K\ge f$, we evaluate $m_R$ on an initial sparse set of feasible ranks up to $K$. Let $B$ be the smallest value of $Rm_R$ found so far. Two bounds allow us
to exclude further ranks.

First, the minimum feasible cap $m_R$ is nonincreasing with the allowed support size. For two evaluated ranks $L<U$, every intermediate rank satisfies
\begin{equation}
Rm_R\ge (L+1)m_U,\qquad L<R<U.
\label{eq:ss-interval-bound}
\end{equation}
If this lower bound exceeds $B$, no rank in the interval can improve the result. Otherwise we evaluate the midpoint and test the two resulting intervals. Second, a rank $R>K$ could still give a smaller product by allowing a sufficiently small cap. To bound this possibility, consider the first $L$ indices. Their original mass is $P_L=\sum_{i=1}^{L}\lambda_i$, and their trial mass is $M_L=\sum_{i=1}^{L}\mu_i$. Applying Cauchy--Schwarz separately to these indices and their complement gives $\sqrt f\le\sqrt{P_LM_L}+\sqrt{(1-P_L)(1-M_L)}$ for every fidelity-feasible $\mu$. Solving this inequality for the minimum possible $M_L$ and using $M_L\le L\max_i\mu_i$ yields the global
cap lower bound
\begin{equation}
m_{\mathrm{LB}}
=
\max_{1\le L\le\min(K,L_{\max})}
\frac{
\left[\sqrt{P_Lf}-\sqrt{(1-P_L)(1-f)}\right]_+^2
}{L},
\qquad [x]_+=\max\{0,x\}.
\label{eq:ss-global-cap-bound}
\end{equation}
Here $L_{\max}$ limits the number of leading sets used to compute the bound. Every unseen rank $R>K$ has objective $Rm_R\ge(K+1)m_{\mathrm{LB}}$. If this exceeds $B$, the current
minimum is global. Otherwise we double $K$ and repeat until the remaining ranks are excluded or the full spectrum has been evaluated.
\end{document}